\documentclass[a4paper,10pt,twoside]{cpc-hepnp}

\usepackage{multicol}
\usepackage{graphicx}
\usepackage{booktabs}
\usepackage{amssymb,bm,mathrsfs,bbm,amscd}
\usepackage[tbtags]{amsmath}
\usepackage{lastpage}

\begin{document}


\footnotetext[0]{Received on the 14th March, 2009}

\title{Electromagnetic transition properties of $\Delta \rightarrow N\gamma$ in a hypercentral scheme }

\author{%
      Kaushal Thakkar\email{kaushal2physics@gmail.com}%
\quad Ajay Majethiya \email{ajay.phy@gmail.com}%
\quad P. C. Vinodkumar\email{p.c.vinodkumar@gmail.com}%
\quad
}
\maketitle

\address{%
Department of Applied Physics, S. V. National Institute of Technology, Surat-395 007, INDIA\\
Kalol Institute of Technology and Research Centre, Kalol, INDIA\\
Department of Physics, Sardar Patel University, Vallabh Vidyanagar-388 120, INDIA.\\
}

\begin{abstract}
The electromagnetic transition properties of the decuplet to octet
baryon ($\Delta \rightarrow N\gamma$) is studied within the frame
work of a hypercentral quark model. The confinement potential is
assumed as hypercentral coloumb plus linear potential. The
transition magnetic moment and transition amplitude $f_{M_1}$ for
the $\Delta \rightarrow N\gamma$ are in agreement with other
theoretical predictions. The present result of the radiative decay
width is found to be in excellent agreement with the experimental
values reported by the particle data group over other theoretical
model predictions.
\end{abstract}



\begin{multicols}{2}

\section{Introduction}
The $\Delta$ (1232) and N (939) are the two lowest decuplet and
octet baryon states in the low flavour (u, d) sector. Their
descriptions and properties thus play a very important role in the
understanding of strong interaction. There are only two decay
channels for $\Delta$ baryon, the first one is $\Delta \rightarrow N\pi$
and the second one is $\Delta \rightarrow N\gamma$. The first decay
channel is more dominant (almost 100$\%$) while the branching ratio
for the second one is less than 1$\%$ \cite{PDG2010}. Due to this
very small branching ratio the electromagnetic transition of $\Delta
\rightarrow N\gamma$ has been the subject of intense study starting
from the early nineties to the very recent times
  \cite{D. B. Leinweber,{S. T. Hong},{T. A. Gail},S. Capstick,T. M. Aliev 2006,T. M. Aliev,{G. Ramalho}}.
However, the high precision measurements of the N $\rightarrow$
$\Delta$ transition by means of electromagnetic probes became
possible with the advent of the
   new generation of electron beam facilities such as BATES, LEGS,
   MAMI,
   ELSA  and at the Jefferson Lab. Several experimental programs
   devoted to the study of electromagnetic properties of the
   $\Delta$ have been reported in the past few years \cite{{PDG2010},A. Bosshard 1991,W. M. Yao}.
These experimental efforts provide new incentives for
  theoretical study of these observables.\\

  The electromagnetic transition of the decuplet to octet baryons is very important to
  understand the internal quark structure and their dynamics. The decuplet to
  octet electromagnetic transitions allowed by spin-parity selection rules are the magnetic dipole ($M_1$),
  electric quadrapole ($E_2$) and Coulomb quadrupole ($C_2$) moments.
  The transition can also give essential information about the shape
  of the baryon. When the shape of the baryons is spherically symmetric, then the $E_2$ and
  $C_2$ amplitudes must vanish. However, the experiments show non
  zero though very small contribution, from $E_2$ and $C_2$ over the
  dominant $M_1$ transition. In the rest frame of $\Delta$ the  $\Delta \leftrightarrow
N\gamma$ process is predominantly an $M_1$ transition involving the
spin and isospin flip of a single quark. The quadrupole amplitudes
are only about 1/40 of the dominant magnetic dipole amplitude.
Though there exist many model predictions on the radiative decay
width of $\Delta \rightarrow N\gamma$ transition based on lattice
calculations, light cone QCD, chiral quark model etc, \cite{D. B.
Leinweber, T. M. Aliev 2006, lang Yu, R. Koniuk, R. Bijker 2000, M.
N. Butler} their predictions vary widely with the experimental
values \cite{PDG2010}. In this article we study the N-$\Delta$
system through a phenomenological hypercentral quark model. The
confinement of the three quark system is described through a
hypercentral coloumb plus linear potential. It is expected that the
quarks confinement effect plays a decisive role in the transition
properties of the baryons. So, we define an effective mass to the
confined quarks within the baryon for the parameter free predictions
of the transition properties of $\Delta \rightarrow
N\gamma$.\\

The article is organized as follows. In Section 2 the hypercentral
scheme and a brief introduction of hypercentral coloumb plus linear
potential employed for the present study are described. Section 3
describes the computational details of transition magnetic moments
and transition amplitude of $\Delta\rightarrow N\gamma$
incorporating with and without the effective mass of the bound
quarks. In Section 4 we present the calculation of the  radiative
decay width ($\Gamma_{M_1}$) of $\Delta\rightarrow N\gamma$ channel.
And in Section 5, we discuss our results while comparing with other
theoretical predictions and experimental results.

\section{Hypercentral scheme for baryons}

The most general Jacobi Co-ordinates to describe a three-body system
of unequal masses can be written as \cite{O Portilho 1991}
\begin{equation}\label{}
\vec{\rho}=\sqrt{\frac{m_2m_3}{m(m_2+m_3)}}(\vec{r}_2-\vec{r}_3)
\end{equation}
\begin{equation}\label{}
\vec{\lambda}=\sqrt{\frac{m_1(m_2+m_3)}{mM}}(\vec{r}_1-\frac{m_2\vec{r}_2+m_3\vec{r}_3}{m_2+m_3})
\end{equation}
\begin{equation}\label{}
\vec{R}=\frac{1}{M}(m_1\vec{r}_1+m_2\vec{r}_2+m_3\vec{r}_3)
\end{equation}
where $m_1$, $m_2$, $m_3$ in our case are the constituent quark mass
parameters, M = $m_1+m_2+m_3$ is the center of  mass of the system
and
\begin{equation}\label{}
m=\frac{1}{M}(m_1 m_2+m_2 m_3+m_1 m_3)
\end{equation}
is equivalent to the reduced mass of the system.\\
The total kinetic energy operator can now be expressed as
\begin{equation}\label{eq:404}
T=\frac{P^2_\rho}{2\,m}+\frac{P^2_\lambda}{2\,m}+\frac{P^2_R}{2\,M}\\
\end{equation}\label{}
Introducing the hyper spherical coordinates which are given by the
angles
\begin{equation}\label{}
\Omega_\rho=(\theta_\rho,\phi_\rho)\,\,{;}\,\,
\Omega_\lambda=(\theta_\lambda,\phi_\lambda)
\end{equation}\label{}
together with the hyper radius, $x$ and hyper angle $\xi$ respectively as \cite{M. Ferraris}\\
\begin{equation}\label{}
x=\sqrt{\rho^2+\lambda^2}\,\,{;}\,\,\xi=\arctan\left(\frac{\rho}{\lambda}\right)\\
\end{equation}\label{}
and assuming the translational invariance, the Hamiltonian in the
hyper central model (hcm) can be written as
\begin{equation}\label{8}
H=\frac{P^2_x}{2\,m}+V(x)\\
\end{equation}\label{}
By expressing the interaction potential of the three-body bound
system in terms of the hypercentral coordinate, x enables us to
incorporate not only the two-body interaction but also the
three-body effects. Such three-body effects are desirable in the
study of hadrons since the non-abelian nature of QCD leads to
gluon-gluon couplings which produce three-body forces. In the six
dimensional hyperspherical coordinates, the kinetic energy operator
$\frac{P^2_x}{2\,m}$ of the three-body system can be expressed as
\begin{equation}\label{}
\frac{P^2_x}{2\,m}=\frac{-1}{2\,m}\left(\frac{\partial^2}{\partial\,x^2}+\frac{5}{x}\frac{\partial}{\partial\,x}-\frac{L^2(\Omega_\rho,\Omega_\lambda,\xi)}{x^2}\right)\\
\end{equation}\label{}
where $L^2(\Omega_\rho,\Omega_\lambda,\xi)$ is the quadratic Casimir
operator of the six dimensional rotational group $O(6)$ and its
eigen functions are the hyperspherical harmonics, $Y_{[\gamma]l_\rho
l_\lambda}(\Omega_\rho,\Omega_\lambda,\xi)$ satisfying the
eigenvalue relation
\begin{equation}\label{}
L^2 Y_{[\gamma]l_\rho l_\lambda}(\Omega_\rho,\Omega_\lambda,\xi)=
\gamma(\gamma+4)Y_{[\gamma]l_\rho
l_\lambda}(\Omega_\rho,\Omega_\lambda,\xi)
\end{equation}\label{}
Here, $\gamma$ is the grand angular quantum number and it takes values 0,1,2....\\
As the interaction potential is hypercentral such that the potential
depends only on the hyper radius $x$,  the hyper radial schrodinger
equation which corresponds to the Hamiltonian given by Eqn.(\ref{8}) can
be written as
\begin{equation}\label{eq:401}
\left[\frac{d^2}{dx^2}+\frac{5}{x}\frac{d}{dx}-\frac{\gamma(\gamma+4)}{x^2}\right]\phi_\gamma(x)=-2m[E-V(x)]\,\phi_\gamma(x)\\
\end{equation}\label{}
Following our earlier study on heavy flavour baryons \cite{Bhavin
2008}, for the present study we consider the hyper central potential
$V(x)$ as the hyper colour coulomb plus linear potential form given
by
\begin{equation}\label{eq:410}
V(x)=-\frac{2}{3}\frac{\alpha_s}{x}+\beta x \\
\end{equation}\label{}
Here $\frac{2}{3}$ is the color factor for the baryon, $\beta$
corresponds to the string tension of the confining term and
$\alpha_s$ is the strong running coupling constant.
To account for the spin dependent part of the three-body
interaction, we add a separate spin dependent potential given by
\cite{Bhavin 2008,{Bhavin 2009}}
\begin{equation}\label{20.5}
V_{spin}(x)=-\frac{1}{4}\alpha_s \, \frac{e^\frac{-x}{x_0}} {x
x^{2}_{0}}\sum\limits_{i{<}j}\frac{\vec{\sigma_i} \cdot
\vec{\sigma_j}}{6m_i m_j} \vec{\lambda_i}\cdot \vec{\lambda_j}
\end{equation}\label{15.5}
to the Hamiltonian. Here, $x_0$ is the hyperfine parameter of the model.\\
The six dimensional radial Schrodinger equation described by
Eq.(\ref{eq:401}) has been obtained in the variational scheme with
the hyper coloumb trial radial wave function given by
\cite{Santopinto1998}

\begin{equation}\label{}
\psi_{\omega\gamma}=\left[\frac{(\omega-\gamma)!(2g)^6}{(2\omega+5)(\omega+\gamma+4)!}\right]^\frac{1}{2}(2gx)^\gamma
e^{-gx} L^{2\gamma+4}_{\omega-\gamma}(2gx)
\end{equation}\label{}
The wave function parameter g and hence the  energy eigen value are
obtained by applying virial theorem.\\
The baryon mass is then obtained as
\begin{equation}\label{eq:415}
M_B=\sum\limits_{i}m_i+\langle H \rangle
\end{equation}\label{}
The model parameters listed in Table \ref{tab:1} are fixed using the
experimental center of weight (spin-average) mass and hyper fine
splitting of the $N-\Delta$ ground state. To account for the quark
confinement effect, we define an effective mass to the bound quarks
as \cite{Bhavin 2008}
\begin{equation}\label{eq:417}
m^{eff}_{i}=m_i\left(
1+\frac{<H>}{\sum\limits_{i}m_{i}}\right) \\
\end{equation}
such that mass of the baryon is given by
 $M_B=\sum\limits_{i=1}^{3}m^{eff}_{i}.$
Accordingly, within the baryon the mass of the quarks may get
modified due to its binding interactions with other two quarks.

\begin{table*}
\begin{center}
\caption{The hypercentral quark model parameters and the
experimental masses of N and $\Delta$} \vspace{0.001in}\label{tab:1}
\begin{tabular}{ccccccc}
\hline $m_u$&$m_d$&$\alpha_s$&$\beta$&$x_0$&N&$\Delta$\\
MeV&MeV&&$GeV^2$&$MeV^{-1}$&MeV&MeV\\
\hline 338&338&1.0&4.4&-0.00496&939&1232\\

\hline

\end{tabular}
\end{center}

\end{table*}









\section{Transition magnetic moment and transition amplitude}

The transition magnetic moment correspond to $\Delta \rightarrow
N\gamma$ can be computed in terms of the orbital and spin-flavour wave function
of the constituent quarks as \cite{{Rohit Dhir 2009 },Fayyazuddin}
\begin{equation}\label{eq:17}
\mu_{\Delta \rightarrow N\gamma} =|\left\langle \Delta_{orb}|j_0(\frac{q\,\, x}{2})|N_{orb}\right\rangle|^2\,\, \sum _{i}\langle \Delta_{sf}\mid \mu_{i}\mathbf {\sigma_{iz}} \mid N_{sf}\rangle
\end{equation}
 Here, the first term corresponds to the contribution from the orbital part of the transition, while the second term is related to the spin flavour contribution to the transition magnetic moment. Here, $j_0(\frac{q\,\, x}{2})$ is the spherical Bessel function, $\mu_{i}=\frac{e_{i}}{2m_{i}^*}$ and q is the photon energy.\\

For the transition $\Delta^+\rightarrow p$, the contribution from the spin flavor wave function is given by
\begin{equation}
(\mu_{\Delta^{+}\rightarrow p\gamma})_{sf}=\sum _{i}\langle \Delta^{+}\mid \mu_{i}\mathbf {\sigma_{iz}} \mid p\rangle
\end{equation}

 The spin-flavour wave function is given by \cite{Fayyazuddin}
\begin{eqnarray}
| p  , s_{z}=\frac{1}{2} \rangle=&\frac{1}{\surd 18}[2u^{\uparrow}d^{\downarrow}u^{\uparrow}+2u^{\uparrow}u^{\uparrow}d^{\downarrow}
+2d^{\downarrow}u^{\uparrow}u^{\uparrow}\nonumber\\&-u^{\uparrow}u^{\downarrow}d^\uparrow-u^{\uparrow}d^{\uparrow}u^{\downarrow}-
u^{\downarrow}d^{\uparrow}u^{\uparrow}\nonumber \\
&-d^{\uparrow}u^{\downarrow}u^{\uparrow}-d^{\uparrow}u^{\uparrow}u^{\downarrow}-u^{\downarrow}u^{\uparrow}d^{\uparrow}]
\end{eqnarray}

and for $\Delta^+$ state there are two possibilities to write down the spin-flavour wave function, 1) $|\Delta^+, S_z=\frac{1}{2}\rangle$ and 2)  $|\Delta^+, S_z=\frac{3}{2}\rangle$. The spin-flavour wave function of these two states can be written as

\begin{eqnarray}
| \Delta^{+}  , s_{z}= \frac{1}{2} \rangle=&\frac{1}{3}[u^{\uparrow}u^{\uparrow}d^{\downarrow}+u^{\uparrow}d^{\uparrow}u^{\downarrow}+d^{\uparrow}u^{\uparrow}u^{\downarrow}\nonumber\\&
+u^{\uparrow}u^{\downarrow}d^{\uparrow}+u^{\uparrow}d^{\downarrow}u^{\uparrow}+d^{\uparrow}u^{\downarrow}u^{\uparrow}\nonumber\\&
+u^{\downarrow}u^{\uparrow}d^{\uparrow}+u^{\downarrow}d^{\uparrow}u^{\uparrow}+d^{\downarrow}u^{\uparrow}u^{\uparrow}]
\end{eqnarray}

\begin{equation}
| \Delta^{+}  , s_{z}= \frac{3}{2} \rangle=\frac{1}{\surd3}(u^{\uparrow}u^{\uparrow}d^{\uparrow}+u^{\uparrow}d^{\uparrow}u^{\uparrow}+d^{\uparrow}u^{\uparrow}u^{\uparrow})
\end{equation}

All there spin-flavour wave functions of p and $\Delta^{+} (S_z=\frac{1}{2},\frac{3}{2})$ are obviously orthogonal to each other. However, the transition matrix element as given by Eqn(17) can provide non zero contribution coming only from  $\langle \Delta^{+}, S_z=\frac{1}{2}| \mu_{i} \mathbf{\sigma_{iz}}|p, S_z=\frac{1}{2} \rangle $ with all other combinations leading to zero.The resulting transition magnetic moment is obtained as

\begin{eqnarray}
\langle \Delta^{+}_{sf}| \mu_{i}\mathbf{\sigma_{iz}}| p_{sf} \rangle=&\frac{1}{3\surd 18}[2(\mu_{u}-\mu_{d}+\mu_{u})\langle u^{\uparrow}d^{\downarrow}u^{\uparrow}|u^{\uparrow}d^{\downarrow}u^{\uparrow}\rangle \nonumber\\&+2(\mu_{u}+\mu_{u}-\mu_{d}) \langle u^{\uparrow}u^{\uparrow}d^{\downarrow}|u^{\uparrow}u^{\uparrow}d^{\downarrow} \rangle+  \nonumber\\
&2(-\mu_{d}+\mu_{u}+\mu_{u})\langle d^{\downarrow}u^{\uparrow}u^{\uparrow}| d^{\downarrow}u^{\uparrow}u^{\uparrow} \rangle - \nonumber\\&(\mu_{u}-\mu_{u}+\mu_{d}) \langle u^{\uparrow}u^{\downarrow}d^{\uparrow}|u^{\uparrow}u^{\downarrow}d^{\uparrow} \rangle- \nonumber\\
&(\mu_{u}+\mu_{d}-\mu_{u})\langle u^{\uparrow}d^{\uparrow}u^{\downarrow}|u^{\uparrow}d^{\uparrow}u^{\downarrow} \rangle - \nonumber\\& (-\mu_{u}+\mu_{d}-\mu_{u}) \langle u^{\downarrow}d^{\uparrow}u^{\uparrow}|u^{\downarrow}d^{\uparrow}u^{\uparrow}\rangle -\nonumber\\ &(\mu_{d}-\mu_{u}+\mu_{u})\langle d^{\uparrow}u^{\downarrow}u^{\uparrow}|d^{\uparrow}u^{\downarrow}u^{\uparrow}\rangle  - \nonumber\\& (\mu_{d}+\mu_{u}-\mu_{u}) \langle d^{\uparrow}u^{\uparrow}u^{\downarrow}|d^{\uparrow}u^{\uparrow}u^{\downarrow} \rangle-\nonumber\\
&(-\mu_{u}+\mu_{u}+\mu_{d})\langle u^{\downarrow}u^{\uparrow}d^{\uparrow}|u^{\downarrow}u^{\uparrow}d^{\uparrow}\rangle]
\end{eqnarray}

\begin{eqnarray}
(\mu_{\Delta^{+}\rightarrow p\gamma})_{sf}&=\frac{1}{3\surd18}[6(2\mu_{u}-\mu_{d})-6\mu_{d}]\nonumber\\
&=\frac{2\surd2}{3}[\mu_{u}-\mu_{d}]
\end{eqnarray}

Similarly we can find out the transition magnetic moment of

\begin{eqnarray}
(\mu_{\Delta^{0}\rightarrow n\gamma})_{sf}=\frac{2\surd2}{3}[\mu_{d}-\mu_{u}]
\end{eqnarray}

As we are not differentiating the different charge states of Delta and the nucleon states (p , n),  we express the transition magnetic moment of $\Delta^+ \rightarrow N\gamma$ in terms of its magnitude only. Finally Eq(\ref{eq:17}) can be reduced into
\begin{equation}
|\mu_{\Delta\rightarrow N\gamma}|=|\left\langle \Delta_{orb}|j_0(\frac{q\,\, x}{2})|N_{orb}\right\rangle|^2\,\,\,\frac{2\surd2}{3}[\mu_{u}-\mu_{d}]
\end{equation}

In Eq(\ref{eq:17}) $e_{i}$ and $\sigma_{iz}$ represents the charge and the spin projection of
the quark constituting the baryonic state. While,
$m_i^*$ is equivalent to the effective mass of the bound quarks of
the N - $\Delta$ system. It is defined in terms of the respective
mass of the bound quarks constituting the N and $\Delta$ states as
\cite{Rohit Dhir 2009 }
\begin{equation}
m_{i}^{eff} = \sqrt{m^{eff}_{i(\Delta)} m^{eff}_{i(N)}}
\end{equation}
Using the spin flavour wave function of the N (octet) and $\Delta$
(decuplet) states \cite{Fayyazuddin}, the transition magnetic moment is
computed with (WEM) and without (WOM) considering the quark
confinment effect through the effective mass of the bound quarks.
The transition amplitude in terms of the transition magnetic moment
can now be computed from the relation
\begin{equation}\label{eq:20}
f_{M_1}^2=\frac{\pi\alpha}{M_N^2}\,\,\,|q|\,\,\, |\mu_{\Delta
\rightarrow N\gamma}|^2
\end{equation}
where, $\alpha$ is the electromagnetic fine structure constant,
$M_N$ is the Nucleon mass, $\mu_{\Delta \rightarrow N}$ is the
radiative transition magnetic moments, q is the photon energy and in
the non-relativistic case, photon energy is given by $M_{\Delta} -
M_{N}$. The present values of transition magnetic moment and
transition ammplitude are listed along with other model predictions
and with the experimental value in Table \ref{tab:5}.\\
\section{Radiative decay}
The radiative decays of baryons provide much better understanding of
the underlying structure of baryons and the dependence on the
constituent quark mass. Though the nonrelativistic model of Isgur
and Karl successfully predicted the electromagnetic properties of
the low lying octet baryons, it fails to provide a good
description of the radiative decay of the decuplet baryons
\cite{Isgur1978}. Thus, the successful prediction of the
electromagnetic properties of octet baryons as well as the decuplet
baryons become detrimental to any phenomenological attempts. The
radiative decay width of the baryon is related to the transition
magnetic moment as \cite{A 2008}
\begin{equation}
\Gamma_{M_1} = {q^3}\,\,  \frac{2}{2J+1}\,\,
\frac{\alpha}{M_N^2}\,\,
 |\mu_{\Delta \rightarrow N}|^2
\end{equation}\label{}
where, J is the total angular momentum quantum number of the
decaying baryon. Thus, in terms of the transition amplitude given by
Eq(\ref{eq:20}), we write the decay width corresponding to
$\Delta\rightarrow N\gamma$ as
\begin{equation}
\Gamma_{M_1}=\frac{q^2}{2\pi}\,\,f_{M_1}^2
\end{equation}
The computed values of radiative decay width and the branching ratio
$\frac{\Gamma_{M_1}}{\Gamma}$ are listed in Table \ref{tab:6}. For
the branching ratio we have used the total decay width of $\Delta$
reported by PDG (2010) \cite{PDG2010}.

\begin{table*}
\begin{center}
\caption{Transition Magnetic Moments ($\mu_{\Delta\rightarrow N}$)
in $\mu_N$ and Transition Amplitude $f_{M1}$ in
$GeV^{-\frac{1}{2}}$} \label{tab:5}
\begin{tabular}{lcccccccc}
\hline {\textbf{Decay Mode}}& \multicolumn{5}{c}{\textbf{\underline
{ Transition Magnetic Moments
($\mu_{N}$)}}}&\multicolumn{3}{c}{\textbf{\underline{Transition
Amplitude}}}
 \\

&Expression&WEM&W0M&others& Expt. \cite{A. Bosshard 1991}&WEM&WOM&Others\\
 \hline

$\Delta\rightarrow N\gamma$&$\frac{2\sqrt{2}}{3}(\mu_{u}-\mu_{d})$&2.6199&2.4699&2.46 \cite{D. B. Leinweber}&3.23$\pm 0.1$&0.2299&0.2199&0.23\cite{D. B. Leinweber} \\
&&&&2.76 \cite{S. T. Hong}\\
&&&&2.50 \cite{T. M. Aliev 2006} \\
&&&&2.48 \cite{Rohit Dhir 2009 }\\
&&&&2.47 \cite{lang Yu} \\


\hline

\end{tabular}\\
WEM - With effective mass,
WOM - Without effective mass\\
\end{center}

\end{table*}

\begin{table*}
\begin{center}
\caption{Radiative Decay Widths ($\Gamma_{M_1}$ in MeV) and
Branching Ratio} \label{tab:6}
\begin{tabular}{lcccccccc}
\hline {\textbf{Decay
Mode}}&\multicolumn{4}{c}{\textbf{\underline{Radiative Decay
Width($\Gamma_{M_1}$) in MeV  }}}
&\multicolumn{4}{c}{\textbf{\underline{Branching Ratio ($\frac{\Gamma_{M_1}}{\Gamma}$) in $\%$}}} \\
&WEM&WOM&Others&Expt. &Symbol&WEM&WOM & Expt. \cite{PDG2010}\\

 \hline

$\Delta\rightarrow N\gamma$&0.7139&0.6389&0.430 \cite{D. B. Leinweber}&0.61-0.71 \cite{PDG2010}&$\frac{\Gamma_{M_1}}{\Gamma(\Delta)}$&0.6049&0.5409&0.52-0.60\\
&&&0.900 \cite{T. M. Aliev 2006}&&&&\\
&&&0.334 \cite{lang Yu}&&&\\
&&&0.36 \cite{R. Koniuk}&&&&\\
&&&0.343 \cite{R. Bijker 2000}&&&&\\
&&&0.67-0.79 \cite{M. N. Butler}&&&&\\
\hline
\end{tabular}
\cite{D. B. Leinweber}-Lattice, \cite{T. M. Aliev 2006}-LCQCD,
\cite{lang Yu}-$\chi$QM, \cite{R. Koniuk}-NRQM, \cite{R. Bijker
2000}-Algebraic model, \cite{M. N. Butler}-HB$\chi$PT
\end{center}

\end{table*}



\section{Results and discussion}
The properties of N, $\Delta$ system are studied within the frame
work of a non-relativistic hypercentral quark model. After fixing
the model parameters using the ground state masses of N and $\Delta$
states, the electromagnetic transition properties are being computed
without any additional parameter. The transition magnetic moment and
the transition amplitude for $\Delta\rightarrow N\gamma$ obtained in
the present study are in good agreement with the lattice result
\cite{D. B. Leinweber} as well as in accordance with other model
predictions. However, all the theoretical predictions are found to
be lower by about 15 to 24$\%$ compared with the experimental value of
$\mu_{\Delta^0\rightarrow n}=3.23$ \cite{A. Bosshard 1991}. The
transition magnetic moments predicted in our model by considering the
confinement effect on the bound quark mass (WEM) and without
considering the confinement effect (WOM) are in good agreement with
other theoretical predictions. The transition amplitude predicted
with effective mass of the bound quarks is in good agreement with
that of the lattice predictions \cite{D. B. Leinweber}. The disagreement
with the theoretical result and experimental value reported by
\cite{A. Bosshard 1991} for the transition magnetic moment of
$\Delta\rightarrow N\gamma$ demands more intense
experimental measurements.\\

In the case of radiative decay width, our results with and without
the bound state effect on the quark mass are in good agreement with
the average range of experimental values (0.61-0.71 MeV) reported by
PDG (2010), while other theoretical values except that by HB$\chi$PT
\cite{M. N. Butler} are widely off by 35 to 50 $\%$. Thus we
conclude here that the hypercentral model with the colour coloumb
plus linear form of the three quarks interactions within baryon is
one of the successful schemes that describes the electromagnetic
properties of the decuplet ($\Delta$) - octet (N) transition. We
look forward to extending the scheme for all the octet and decuplet
baryons in the (u, d, s) sector.

\end{multicols}

\vspace{-1mm}
\centerline{\rule{80mm}{0.1pt}}
\vspace{2mm}

\begin{multicols}{2}

\end{multicols}

\clearpage

\end{document}